# Analysis of the first confirmed superoutburst of V337 Cygni in 2006 May

David Boyd, Tom Krajci, Jeremy Shears & Gary Poyner

During 2006 May the first confirmed superoutburst of the dwarf nova V337 Cygni was observed using unfiltered CCD photometry. The outburst reached a peak magnitude of 15.2 on May 22 and lasted 13 days before the object returned to quiescence at around magnitude 20. Time-series photometry revealed superhumps with a period of 0.07012 ± 0.00033d and amplitude 0.1 magnitude, thereby establishing for the first time its UGSU classification. Astrometry of V337 Cyg in outburst determined an accurate position and enabled the identity of the star in quiescence to be confirmed.

## Introduction

Dwarf novae are a class of cataclysmic variable stars in which a cool main sequence secondary star loses mass to a white dwarf primary. Material from the secondary falls through the inner Lagrangian point and, because it carries substantial angular momentum, does not settle on the primary immediately but forms an accretion disc. From time to time, as material builds up in the disc, thermal instability drives the disc into a hotter, brighter state causing an outburst in which the star brightens by several magnitudes. Dwarf novae of the SU UMa family (UGSU) occasionally exhibit superoutbursts which last several times longer than normal outbursts and may be up to a magnitude brighter. Superoutbursts are completely diagnostic of the UGSU class. During a superoutburst the light curve of a UGSU star is characterised by superhumps. These are modulations which are a few percent longer than the orbital period and are thought to be caused by precession of the accretion disc.[1]

## History of V337 Cyg

V337 Cygni is listed in both the General Catalogue of Variable Stars (GCVS)[2] and the Catalog and Atlas of Cataclysmic Variables – Archival Edition[3] as a UG-type dwarf nova with an unknown orbital period. It was discovered in 1928 by Walter Baade during his survey of variable stars in Sagitta and Cygnus and reported with an amplitude range of 14.2 to >16.4 $m_{pg}$.[4] Brian Skiff correlated Baade's variable AN 101.1928 Cyg with V337 Cyg.[5] However, it is worth noting that the positions of the variable given by both Baade (when precessed to epoch J2000) and Skiff correspond to the position of a nearby 15th magnitude star (marked 151 on the AAVSO chart, see below). There has been uncertainty about the correct identity of V337 Cyg at quiescence which we resolve in this paper. In passing, it may be remarked that the stars reported in Baade's survey are of some historical interest, because this work led to his later recognition of the two stellar populations of the Milky Way.

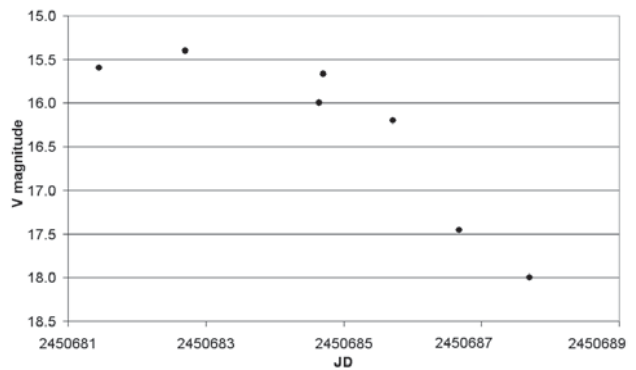

**Figure 1.** Light curve of the 1997 August outburst. V-band photometry from the AAVSO International Database.

In 1994 Jan Manek at the Stefanik Observatory, Czech Republic, examined the plate collection at the Sonneberg Observatory in an attempt to recover Baade's variables in outburst. He found at least three outbursts of V337 Cyg, usually less than five days long and reaching about magnitude 15 at maximum on blue plates. At minimum it was below the plate limit of 18 $m_{pg}$.[6] Manek later reported three more outbursts on archive plates, and noted a possible recurrence period of 30 or 32 days.[7]

The first outburst of V337 Cyg detected as it happened was reported by Manek at 15.6V on 1997 Aug 20.95 using the 0.65m telescope at the Stefanik Observatory.[8] The lightcurve of the outburst (Figure 1), based on V-band photometric measurements by four observers in the American Association of Variable Star Observers (AAVSO) International Database,[9] indicates a maximum brightness of 15.4V. The last positive observation at 18.0V was six days after discovery on 1997 August 27.2. No time-series photometry appears to have been carried out during the outburst to look for superhumps, but the duration and peak magnitude appear to be compatible with the archival outbursts.

We also searched the AAVSO database for evidence of other outbursts and discovered observations by Lasse Jensen at 16.2C and 16.0C on 1996 July 16 and 17 respectively. Patrick Schmeer reported V337 Cyg at 17.0CR on 2003 July 7.[10] Because of limited coverage it is not clear whether these were full dwarf nova outbursts or examples of the brief, less bright, outbursts reported in some systems such as V1316 Cyg.[11] That no other outbursts have been reported calls into question Manek's assertion of a 30 or 32 day outburst cycle. Either many outbursts have been missed or the outburst cycle is longer than Manek claimed.

## Observations

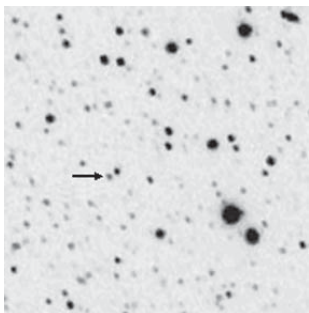

**Figure 2.** V337 Cyg in outburst at 16.0C on 2006 May 20.956. 5'×5' with south at top *(Jeremy Shears)*.

V337 Cyg was added to the British Astronomical Association Variable Star Section (BAAVSS) Recurrent Objects Programme in 1994.[12] This programme was set up as a joint project between the BAAVSS and *The Astronomer* magazine specifically to monitor poorly studied eruptive stars of various types. We have monitored V337 Cyg since 2005 April with a limiting magnitude of around 17.0C depending on conditions. Over 14 months the field of V337 Cyg was imaged on 84 nights. V337 Cyg was positively detected on only one occasion at 16.1C on 2005 May 29.

The superoutburst reported here was first detected by Shears[13] on 2006 May 20.956 at 16.0C (Figure 2). Over the next 21 days, V337 Cyg was observed on 26 occasions using unfiltered CCD photometry, producing 23 individual magnitude measurements and 3 time-series runs totalling 13.8 hours. Table 1 summarises the instrumentation used and Table 2 contains a log of the time-series runs. Exposures were of 60s duration increasing to 120s at the end of the outburst and were made unfiltered to maximise signal to noise ratio. Images of the object below magnitude 18 were stacked before measurement, again to increase signal to noise ratio. CCD images were dark-subtracted and flat-fielded before being measured using differential aperture photometry with the AIP4WIN package.[14] The AAVSO has produced charts for V337 Cyg,[9] and comparison stars from their f-scale chart (the central section of which is reproduced in Figure 3) were used in reducing the observations in this paper. We used comparison star photometry by Henden given in Sumner.[15] Close proximity of the 15th magnitude star labelled 151 on the AAVSO f-chart necessitated care to ensure that light from this star did not contaminate either the photometric aperture for V337 Cyg or measurements of sky background. The difficulty in doing this when V337 Cyg was faint is reflected in larger errors. Use of the Bradford Robotic Telescope on Tenerife enabled us to cover the early stages of the outburst when weather conditions in most of mainland Europe and North America were poor.

**Table 1. Instrumentation used**

| Observer | Instrumentation |
|---|---|
| JS | 0.1m fluorite refractor + SXV-M7 CCD camera |
| GP | 0.36m SCT + FLI MaxCam ME2 CCD camera (Bradford Robotic Telescope on Tenerife) |
| DB | 0.36m SCT + SXV-H9 CCD camera |
| T K | 0.28m SCT + ST-7E CCD camera |

**Table 2. Log of time-series observations**

| Start time (JD) | Duration (hr) | No of images | Mean mag (unfiltered) | Observer |
|---|---|---|---|---|
| 2453886.422 | 2.8 | 149 | 15.84 | DB |
| 2453886.675 | 7.1 | 374 | 16.03 | T K |
| 2453888.428 | 3.9 | 202 | 16.20 | DB |

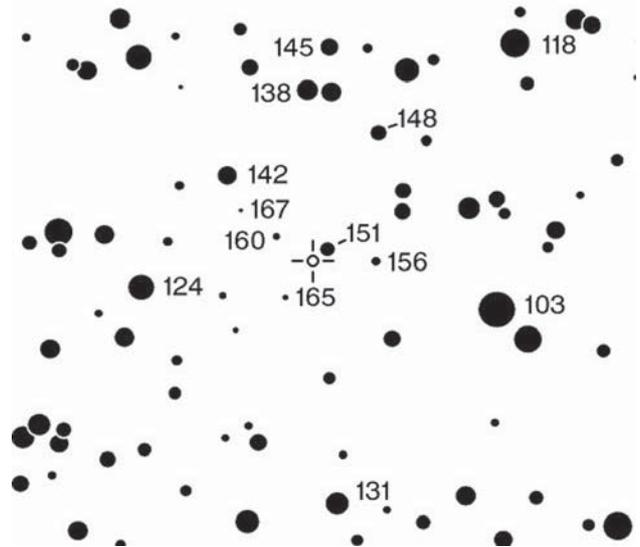

**Figure 3.** Segment of AAVSO f-scale chart of V337 Cyg. (Magnitudes have decimal point removed.) South at top.

Figure 4 shows the lightcurve of the outburst. From maximum light at 15.2C on May 22, the outburst declined at an average rate of 0.09 mag per day for 11 days and then dropped sharply. Overall the outburst lasted 13 days which is two to three times longer than the outbursts seen by Manek, and although the peak magnitude was not obviously brighter, this behaviour is consistent with a superoutburst. On June 27, 38 days after the outburst was first detected, a 25 min stacked exposure recorded the object at magnitude 20, presumably close to its quiescent level, indicating the outburst amplitude was approximately 5 magnitudes.

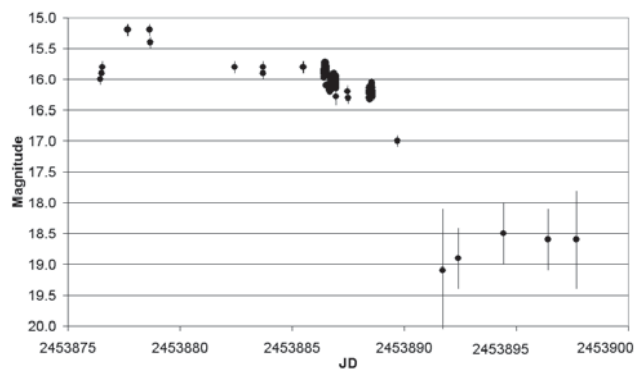

**Figure 4.** Unfiltered light curve of the 2006 May outburst.

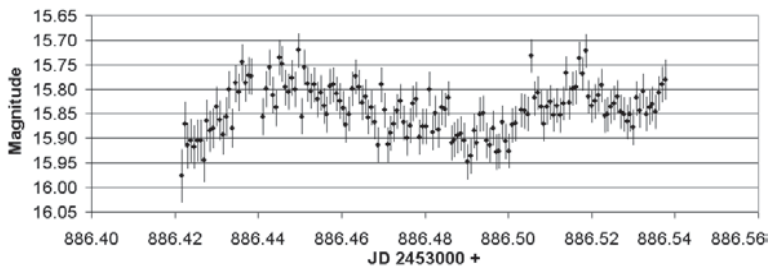

**Figure 5.** First observation of superhumps on 2006 May 30 *(David Boyd)*.

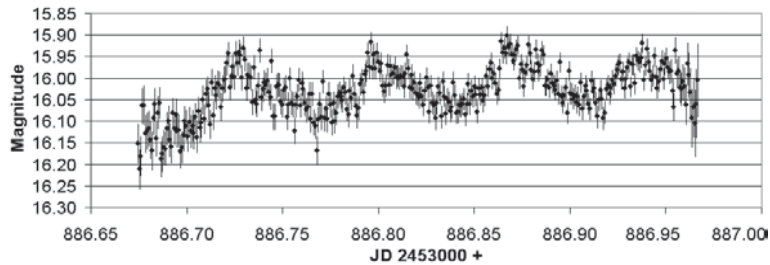

**Figure 6.** Confirmation of superhumps on 2006 May 31 *(Tom Krajci)*.

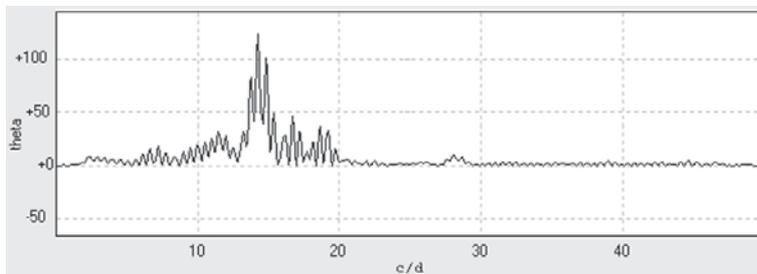

**Figure 7.** Power spectrum of combined time-series data.

## Detection of superhumps

The time-series photometry data obtained on May 30 (Figure 5) provided the first detection of superhumps in V337 Cyg, thereby confirming this object to be a member of the UGSU class of dwarf novae. The peak-to-peak superhump amplitude was 0.1 magnitude. Notification of this discovery was quickly disseminated by e-mail on the baavss-alert and cvnet-outburst distribution lists. Follow-up observations were begun 3 hours later in New Mexico and, with the advantage of a longer period of darkness, we obtained a 7-hour photometry run; a good example of the practical benefit of rapid global communication. This longer run (Figure 6) confirmed the presence of superhumps and improved the period determination. Further time-series data on June 1 showed the superhumps continuing but with slightly diminished amplitude. Poor weather prevented any further time-series observations.

## Superhump and orbital periods

After subtracting the mean and linear trend from each of the time-series light curves, they were combined and period analysed using the CLEANest algorithm in Peranso.[16] Figure 7 shows the resulting power spectrum. The strongest signal arises from the superhumps and gives the superhump frequency as $14.26 \pm 0.07$ c/d, and the corresponding period $P_{sh} = 0.07012 \pm 0.00033$d. The uncertainty is calculated using the Schwarzenberg–Czerny method.[17]

There are various empirical and theoretical relationships between the orbital period $P_{orb}$, the period excess $\varepsilon = (P_{sh} - P_{orb}) / P_{orb}$, and the mass ratio of the components q, based on observation and on modelling both dynamical precession and disc pressure effects in cataclysmic variables.[18,19] From the observed distribution of $\varepsilon$ vs $P_{orb}$, and assuming V337 Cyg is not abnormal, it is possible to estimate that $P_{orb} \sim 0.068$d, $\varepsilon \sim 0.037$ and $q \sim 0.170$. These values are consistent with results of the various empirical and theoretical relationships assuming a primary mass of ~0.7 solar masses. Radial velocity measurements or photometry at quiescence are required to measure $P_{orb}$.

Removing $P_{sh}$ from the power spectrum leaves only weak signals, none of which have any significant relationship to the superhump or orbital periods. Folding the combined time-series data on $P_{sh}$ gives the phase diagram in Figure 8. The phase is coherent across the three time-series runs indicating no detection of the superhump phase change which is sometimes seen in UGSU systems, although the time-series observations cover a relatively short part of the outburst.

## Position and identity of V337 Cyg

There has been confusion in the past about the co-ordinates and identity of V337 Cyg. Various positions have been given and two faint objects are located within the identification circle in Downes *et al*.[3] Astrometry during the outburst using *Astrometrica*[20] and the USNO *CCD Astrograph Catalog* release 2 gives the position of V337 Cyg as RA 19h 59m 52.93s ± 0.10s, Dec +39° 13' 59.94" ± 0.10" (J2000).

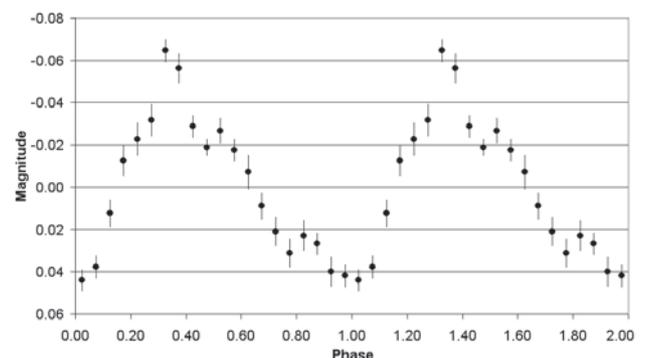

**Figure 8.** Phase diagram of time-series data folded on $P_{sh}$.

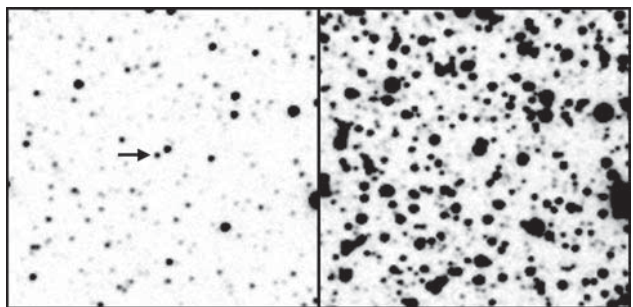

**Figure 9.** V337 Cyg in outburst on 2006 May 30 *(left)* and quiescence June 27 *(right)*. The central star is 151 on the AAVSO f-chart, exposures 1 min (left) and 25 min (right), fields 4'×4' with south at top *(David Boyd)*.

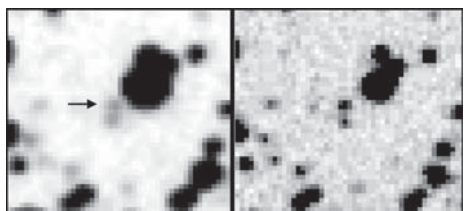

**Figure 10.** Identity of V337 Cyg in quiescence. V337 Cyg *(left, indicated)* is the more southerly of two faint stars NW of star 151 in this enlarged detail from Figure 9; *(right)* segment from the POSS-II blue plate, south at top.

Figure 9 shows two images of the field around V337 Cyg, during outburst on May 30 (1 min exposure) and at quiescence on June 27 (25 min stacked exposure). Figure 10 shows an enlargement of the quiescent image together with the corresponding area on the Digitised Sky Survey POSS-II blue plate, in both of which two faint stars approximately 4 arcsec apart can be seen NW of star 151 on the AAVSO chart. Astrometry of the 25 min stacked exposure gives the position of the more southerly of these two stars as RA 19h 59m 52.96s ± 0.17s, Dec +39° 13' 59.89" ± 0.15" (J2000), in good agreement with the position of the object in outburst, confirming that this star is V337 Cyg.

## Acknowledgments

The authors acknowledge with thanks access to observations of the 1997 August outburst of V337 Cyg from the AAVSO International Database contributed by observers worldwide. We thank Dr Arne Henden, Director of the AAVSO, for permission to reproduce the AAVSO chart of V337 Cyg. We acknowledge use of images from the Digitised Sky Surveys. We gratefully acknowledge our use of the Bradford Robotic Telescope operated by the Department of Cybernetics, University of Bradford, UK, located on Tenerife. Finally we thank Dr Chris Lloyd (OU) for many helpful comments on drafts of the paper, and the referees for their constructive suggestions.

**Addresses: DB:** 5 Silver Lane, West Challow, Wantage, Oxon, OX12 9TX, UK [drsboyd@dsl.pipex.com]
**TK:** CBA New Mexico, PO Box 1351 Cloudcroft, New Mexico 88317, USA [tom_krajci@tularosa.net]
**JS:** 'Pemberton', School Lane, Bunbury, Tarporley, Cheshire, CW6 9NR, UK [bunburyobservatory@hotmail.com]
**GP:** 67 Ellerton Road, Kingstanding, Birmingham, B44 0QE, UK [garypoyner@blueyonder.co.uk]